\documentclass[12pt]{article}

\usepackage[super,comma]{natbib}

\usepackage[pdftex]{graphicx}
\usepackage{latexsym,amsmath}
\usepackage{color}

\usepackage[format=plain,labelfont=bf,up]{caption}
\usepackage{hyperref}

\newcommand{\av}[1]{\langle #1 \rangle}
\newcommand{\fluc}[1]{\langle {#1}^2\rangle}

\usepackage{color}

\begin{document}

\title{Competing activation mechanisms in epidemics on networks}

\author{Claudio Castellano$^{1,2}$ and Romualdo Pastor-Satorras$^{3,*}$}

\maketitle
\begin{center}
\small  
$^1$Istituto dei Sistemi Complessi (ISC-CNR),
Via dei Taurini 19, I-00185 Roma, Italy\\
$^2$Dipartimento di Fisica, ``Sapienza''
Universit\`a di Roma, P.le A. Moro 2, I-00185 Roma, Italy\\
$^3$Departament de F\'\i sica i Enginyeria Nuclear,
  Universitat Polit\`ecnica de Catalunya, Campus Nord B4, 08034
  Barcelona, Spain\\
$^*$ Corresponding author
\end{center}

\begin{abstract} 
   In contrast to previous common wisdom that epidemic activity in
  heterogeneous networks is dominated by the hubs with the largest
  number of connections, recent research has pointed out the role that
  the innermost, dense core of the network plays in sustaining
  epidemic processes.  Here we show that the mechanism responsible of
  spreading depends on the nature of the process. Epidemics with a
  transient state are boosted by the innermost core. Contrarily,
  epidemics allowing a steady state present a dual scenario, where
  either the hub independently sustains activity and propagates it to
  the rest of the system, or, alternatively, the innermost network
  core collectively turns into the active state, maintaining it
  globally.  In uncorrelated networks the former mechanism dominates
  if the degree distribution decays with an exponent larger than
  $5/2$, and the latter otherwise.  Topological correlations, rife in
  real networks, may perturb this picture, mixing the role of both
  mechanisms.
\end{abstract}

\clearpage

The discernment of the mechanisms that contrive to activate spreading
processes on heterogeneous substrates is a pivotal issue, with
practical applications ranging from the containment of epidemic
outbreaks~\cite{PhysRevLett.91.247901} to the viral spreading of
rumors and beliefs~\cite{Leskovec:2007:DVM:1232722.1232727,Daley64}.
The interest on the effects of heterogeneity has been brought about by
the observation that social contact networks (the natural substrate
for most human epidemic processes) are generally strongly
heterogeneous~\cite{barabasi02,Dorogovtsev:2002,newman2003saf},
observation that has led to the introduction of complex network theory
in the quantitative analysis of epidemic
spreading~\cite{keeling05:_networ}. In this context, the nature of the
activation mechanisms translates on simple epidemic models
\cite{anderson92} in setting the epidemic threshold $\lambda_c$ for
some rate of infection $\lambda$ (the spreading rate), separating a
phase in which the spreading affects a finite fraction of the
population from a state in which only a vanishingly small fraction is
hit. The research effort is thus focused on a twofold objective: The
identification of the activation mechanisms as a function of the
network topology, and the determination of the functional form of the
epidemic threshold.

For the sake of concreteness, we focus our discussion on the simplest
models of disease spreading, namely the
susceptible-infected-susceptible (SIS) and the
susceptible-infected-recovered (SIR) models, leading, respectively, to
a steady endemic state or to transient outbreaks affecting a given
fraction of the population~\cite{anderson92} (see Methods). On a
network substrate---statistically described, at the simplest level, by
its degree distribution $P(q)$, defined as the probability that a
randomly chosen individual (vertex) is connected to $q$ other
individuals~\cite{barabasi02}---the application of a heterogeneous
mean-field (HMF) approach \cite{barratbook} assuming no topological
correlations \cite{serrano07:_correl} and neglecting dynamical
correlations, yields epidemic thresholds inversely proportional to the
second moment of the degree distribution, $\av{q^2}$
\cite{marianproc,Lloyd18052001,PhysRevE.66.016128}. Since most natural
networks have a degree distribution scaling as~\cite{barabasi02} $P(q)
\sim q^{-\gamma}$, $\av{q^2}$ takes the form, in the continuous degree
approximation, $\av{q} \sim q_{max}^{3 -\gamma}$, where $q_{max}$ is
the maximum degree in the network \cite{Dorogovtsev:2002}. The second
moment therefore diverges in the infinite network size limit
(i.e. when $q_{max}\to\infty$) for $\gamma \leq 3$, leading to a
vanishing epidemic threshold, i.e. any disease can invade the system,
whatever its infection rate \cite{pv01a,PhysRevLett.90.028701}. For
$\gamma > 3$, on the other hand, HMF predicts a finite threshold. This
result has usually been interpreted in terms of the leading role of
the hubs (the vertices with largest degree in the network) as the
elements sustaining the epidemic activity in the network, whenever
they have a sufficiently large degree to make the second moment
$\av{q^2}$ diverge (i.e. when $\gamma \leq 3$) \cite{barratbook}.

More refined approaches than HMF, incorporating the effects of the
quenched topological structure of the network, but still neglecting
dynamical correlations~\cite{Wang03,0295-5075-89-3-38009,Mieghem2011},
predict that the epidemic threshold for SIS is in general set by the
largest eigenvalue $\Lambda_N$ of the adjacency matrix,
i.e. $\lambda_c^E = 1/\Lambda_N$. This finding, combined with the
scaling of $\Lambda_N$ (computed by Chung et al. for a class of finite
graphs with degrees distributed according to a
power-law~\cite{Chung03}), $\Lambda_N \sim
\max[\sqrt{q_{max}},\av{q^2}/\av{q}]$, leads to a
threshold~\cite{PhysRevLett.105.218701} %
\begin{equation} 
	\lambda_c^E \simeq \left \{ \begin{array}{lr} 
	1/ \sqrt{q_{max}} & ~~~~~~~ \ \	\gamma > 5/2 \\
        \frac{\av{q}}{\av{q^2}} & ~~~~~~~ 2< \gamma < 5/2
\end{array} \right. .
\label{together} 
\end{equation} 
Equation~\eqref{together} implies that the epidemic threshold vanishes
in the thermodynamic limit in power-law distributed networks for any
value of $\gamma$, even larger than $3$, as long as $q_{max}$ is a
growing function of the network size $N$, in agreement with previous
results for SIS~\cite{Ganesh05, Chatterjee09, Durrett10}. In this
perspective, it would be the hub, or most connected vertex, the main
responsible of maintaining the epidemic activity and correspondingly
setting the threshold \cite{PhysRevLett.105.218701}.

The relevance of hubs has been however recently called into question
by Kitsak et al.~\cite{kitsak2010}, who pointed out that in some real
networks, the most efficient spreaders are located at the innermost,
dense core of the network, as identified by means of a $k$-core
decomposition~\cite{Seidman1983269} (see Methods and
Figure~\ref{fig:examplenet}). In this alternative view, it is thus the
nucleus of high $k$-core index which sustains epidemic activity,
independently of the degree of the vertices it is composed of.

Inspired by these results, here we analyze in detail the role of the
hub and of the core of the network for the onset of epidemic spreading on
complex topologies. By means of theoretical arguments and extensive
numerical simulations, we are able to show that the leading mechanism
governing the dynamics depends on the network features, in particular
on the strength of the degree heterogeneity, as measured by the degree
exponent $\gamma$. The analysis of real networks allows to determine
additionally the critical role of degree correlations in suppressing
or enhancing the relevant mechanism.  The findings presented in this
work represent an advancement in the understanding of the underlying
mechanisms that control the behavior of epidemic processes on complex
heterogeneous networks. By identifying with precision the set of
vertices ultimately responsible for the epidemic activation, our
results open the path for the formulation of immunization strategies
\cite{PhysRevLett.91.247901,PhysRevE.65.036104} specifically tailored
for each particular network configuration considered. Moreover, our
results can find application in other, more general spreading
processes, such as rumor, behavior or information spreading in
networks \cite{Centola2010Spreading,1367-2630-13-12-123005}, as well
as other dynamical processes ruled by the largest eigenvalue of the
adjacency matrix, such as synchronization phenomena
\cite{Arenas200893}.

\section*{Results}

\subsection*{Activation mechanisms for SIS in uncorrelated networks}

In the case of the SIS model the expression of the threshold
$\lambda_c$ for $\gamma>5/2$ can be understood by considering the
largest hub and its neighbors as a star network of size $q_{max}+1$.
Such a system has, in isolation, a threshold $\lambda_c^{star} =
1/\sqrt{q_{max}}$ and is thus capable, all by itself and independently
of the degree of the rest of the vertices, to propagate the infection
to a finite fraction of the network, leading to a stable endemic state
whenever $\lambda >
\lambda_c^{star}$~\cite{PhysRevLett.105.218701}. It is therefore the
most connected vertex which singlehandedly can keep the epidemic
activity alive, setting in this way the global threshold for activity
in the system. The change for $\gamma<5/2$ in
equation~\eqref{together} is however surprising and hints towards the
possibility of different activation mechanisms for different $\gamma$
values, thus challenging the belief in the preponderant role of hubs,
which has become common wisdom in network science~\cite{barratbook}.
The results of Kitsak \emph{et al.} would fit in place, pointing
towards a preponderant role for the innermost core of the
network. However, while the picture presented by Kitsak et
al.~\cite{kitsak2010} is compelling for the SIR model, the case of the
SIS deserves a closer look.

In order to shed light on this issue, we have performed extensive
numerical simulations of the SIS process on synthetic uncorrelated
scale-free networks with degree distribution $P(q) \sim q^{-\gamma}$,
generated via the uncorrelated configuration model
(UCM)~\cite{ucmmodel} (see Supplementary Information online for more details). 
We have computed the density of infected
vertices in the whole network, and the same density when the dynamics
takes place (in isolation) on the $k$-core of highest index (maximum
$k$-core) and on the star-graph centered around the hub of the
network, with degree $q_{max}$.

In Figure~\ref{fig:UCMdata} we show the evolution of the recorded
densities as a function of time for different values of the spreading
rate $\lambda$, in networks with large and small degree exponents,
namely $\gamma=2.75$ and $\gamma=2.1$. Our results show a remarkable
dependence on the degree exponent: For large $\gamma$, the onset of a
global stationary state takes place for the same values of $\lambda$
for which the star-graph centered around the hub starts to be active,
while the maximum $k$-core remains subcritical, with exponentially
decaying activity. This behavior is a proof of the leading role of the
hub as the main activation mechanism for large $\gamma$. For small
values of $\gamma$, instead, the picture is opposite: For values of
$\lambda$ corresponding to a globally active network, the maximum
$k$-core is in an active state, while the star-graph centered around
the hub is inactive, indicating that now the maximum $k$-core is the
trigger activating the whole system. Two observations are in order:
The maximum $k$-core is the heart of the nucleus of most densely
connected vertices in the network but it does not sharply coincide
with it. Other nodes, belonging to $k$-cores of index slightly
smaller, are also densely connected. The transition in the whole
network is influenced also by these other nodes and therefore only
approximately coincides with the transition of the maximum $k$-core.
This explains why in Fig.~\ref{fig:UCMdata} the whole network is fully
active for $\lambda=0.01$, while the maximum $k$-core is still around
the transition. The second observation is that in uncorrelated
networks the hub usually belongs to the maximum $k$-core. Yet the two
activation mechanisms for epidemics are clearly distinct. In one case
(hub triggered activation) the hub alone is able to sustain activity
in the set of its neighbors and then propagate it to the rest of the
system. In the other (maximum $k$-core triggered activation) the hub
alone is not able to sustain activity: Only the presence of all
densely connected vertices in the $k$-core allows them to collectively
turn into the active state and propagate to the rest of the system.

This change of behavior with the degree exponent, which we have
confirmed for different values of $\gamma$ above and below $5/2$
(see Supplementary Information Figure S1 online), can
be made more physically transparent by linking it analytically with
the different thresholds in equation~\eqref{together}.  To do so, we
estimate the threshold associated to the active maximum $k$-core,
whose index is denoted as $k_S$.  The maximum $k$-core has a degree
distribution which is bounded and narrow, with minimum ($k_S$),
average, and maximum degree scaling with size in the same way (see
Supplementary Information Figure S2 online).  Hence its epidemic threshold is
well approximated by $\lambda_c^K = 1/\av{q} \sim 1/k_S$.  On the
other hand, in Ref.~\cite{PhysRevLett.96.040601} the maximum $k$-core
index $k_S$ was determined as a function of the network topology,
yielding for scale-free networks with $2 < \gamma < 3$
\begin{equation}
  \label{eq:1}
  k_{S} \approx (\gamma-2)(3-\gamma)^{(3-\gamma)/(\gamma-2)} q_{max}
  \left(\frac{q_{min}}{q_{max}}\right)^{(\gamma-2)},
\end{equation}
where $q_{min}$ is the minimum degree.  Introducing this result into
the formula for $\lambda_c^K$ we obtain $\lambda_c^K \sim
q_{max}^{\gamma-3}$.  It is most noteworthy that the scaling behavior
of the maximum $k$-core threshold $\lambda_c^K $ takes the exact same
form as the eigenvalue threshold for $\gamma<5/2$ in
equation~\eqref{together}.  This observation provides a physical
interpretation of the different activation mechanisms and associated
thresholds in uncorrelated scale-free networks: When $\gamma <5/2$,
the epidemic transition is collectively triggered by the vertices in
the innermost core and the threshold is correspondingly given by
$\lambda_c \sim 1/\av{q^2} \sim q_{max}^{\gamma-3}$, as in HMF theory.
On the other hand, for $\gamma > 5/2$, the hub triggers the global
activity, and the threshold is given by $ \lambda_c \simeq
1/\sqrt{q_{max}}$.  An additional inspection of the numerical values
of the different thresholds (see Supplementary Information Table S1 online)
shows that the thresholds computed from the numerical estimation of
the largest eigenvalue of the adjacency matrix are in very good
agreement with the predictions of Eq.~\eqref{together}, perfectly
accounting for the results observed in Figure~\ref{fig:UCMdata}.

\subsection*{The SIR model} 
\label{sub:the_sir_model}


We turn now our
attention to the SIR model, which, contrarily to the steady-state
dynamics of the SIS model, exhibits transient outbreaks
characterized by the number of infected individuals, totaling a finite
fraction of the system only above the epidemic threshold. 
Evidence that the hub
plays no special role in SIR dynamics comes from considering this process on
a star network of size $q_{max}+1$. For an
epidemics starting from a randomly chosen vertex, the average final 
density of infected nodes takes the form (see Methods)
\begin{equation}
  \label{eq:2}
  R = \lambda^2 + \frac{1+2\lambda}{q_{max}}.
\end{equation}
Hence the threshold in a star network, defined by the value of
$\lambda$ above which $R$ takes a given fixed finite value, is a
constant independent of $q_{max}$, in the limit of large $q_{max}$. 
The hub cannot therefore be the ultimate trigger of
global outbreaks in the SIR model, and this role must instead be
played by the maximum $k$-core for any value of $\gamma$, in
accordance with Ref.~\cite{kitsak2010}. Additional support for this
view comes from extending to the SIR case the maximum $k$-core threshold
argument presented for the SIS model.
Approximating again the maximum $k$-core as a narrowly distributed
graph of average degree $\av{q} \sim k_S$, a threshold is obtained
from MF theory of the form $\lambda_c^{SIR,K} \sim 1/k_S$.  Given the
form of $k_S$ in equation~\eqref{eq:1}, this threshold scales in the
large network limit in exactly the same form as the HMF prediction,
namely $\lambda_c^{SIR} = \av{k}/[\fluc{k} - \av{k}]$
\cite{Cohen00,newman02}.  The conclusion is that in the SIR model it
is always the maximum $k$-core which controls epidemic spreading and
sets the threshold to the HMF value.  This picture is substantiated in
Figure~\ref{fig:UCMSIR}, where we consider the SIR model on UCM
networks with different values of $\gamma$, keeping track, as a
function of $\lambda$, of the density of infected individuals in the
global network, in the maximum $k$-core and in the star-graph centered
around the hub. As we can see, the position of the transition to a
finite fraction of infected vertices is closely correlated in the
whole network and the maximum $k$-core, while the behavior of the hub
conforms to the prediction of equation~\eqref{eq:2}.

\subsection*{Effects of correlations}

The scenario discussed so far applies to the case of uncorrelated
networks, where the probability that a random edge is connected to a
vertex of degree $q$ is proportional to $q P(q)$
\cite{Dorogovtsev:2002}. Real networks, however, present in most cases
some level of degree correlations \cite{serrano07:_correl}, as
measured by the Pearson coefficient \cite{PhysRevLett.89.208701} or by
the average degree of the nearest neighbors (ANN) of the vertices of
given degree, $\bar{q}_{nn}(q)$ \cite{alexei}.  In order to ascertain
their effect on the relevant epidemic mechanisms, we have considered
the SIS process on several instances of correlated real networks (see
Methods): An Internet map at the autonomous system (AS) level, the
social network of pretty-good-privacy (PGP), and the network of actors
co-starring in Hollywood movies (Movies).  All these networks have a
degree distribution compatible with a power law, with an exponent
close to $2$ (see Supplementary Information Figure S3 online) and a
range of degree correlations (see Supplementary Information Figure S4
online).  In this case, according to our arguments above and
neglecting correlations, one would expect the transitions to be ruled
by the corresponding maximum $k$-cores.  This fact is confirmed for
the Movies and PGP networks, by the SIS simulations presented in
Figure~\ref{fig:Realdata}, showing that the transition occurs
simultaneously for the maximum $k$-core and the whole system, while
the star-graph centered around the hub remains inactive.  In the case
of the AS network, instead, the picture is surprisingly the opposite,
 and it is apparently the hub the
  responsible of the epidemic transition, see
Figure~\ref{fig:Realdata}.  The situation is still more complex
  when one considers other, larger AS maps (see Supplementary
  Information). In fact, as it turns out from the analysis of
  numerical simulations (see Supplementary Information Figure S5
  online), the general situation in AS maps is that we can reach in
  the network an active, infected state for values of $\lambda$ for
  which both the hub and the maximum $k$-core are apparently
  subcritical. This observation hints towards a mixing of activation
  mechanisms for the particular case of AS networks.
This discrepancy between
the AS and the other networks, confirmed by the inspection of the
values of the different thresholds (see Supplementary Information
Table S2 online) can be attributed to the presence of strong degree
correlations. Measuring them by means of the auxiliary ANN function
$\bar{q}_{nn}$, we can observe that the AS network is strongly
correlated, with $\bar{q}_{nn}$ decaying as $q^{-0.5}$. Moreover,
these correlations are so strong that they do not wash away even after
randomizing the network, as they do in the PGP and Movies networks
(see Supplementary Information Figure S4 online).  Strong disassortative
correlations reduce the interconnections of vertices of high-degree,
suppressing in this way the index $k_S$ of the maximum $k$-core
and reducing the number of vertices that compose it.  This situation,
i.e. a very large hub coupled with a relatively small maximum
$k$-core, leads to a mixing of both mechanisms that does not allow to
make explicit prediction about the most relevant one. Similar or opposite
effects (i.e. strong assortativity enhancing the role of the maximum
$k$-core) can be found in networks generated by means of the
Weber-Porto algorithm~\cite{PhysRevE.76.046111}, a modification of the
configuration model that generates graphs with prescribed degree
distribution and correlations of tunable strength (see Supplementary
Information Figure S6 and Table S2 online).

\section*{Discussion}

The rationalization of the different mechanisms that keep an epidemic
process alive in a heterogeneous substrate turns out to be a more
complex issue than previously believed. In fact, two different subsets
of vertices (either the hub or the innermost core of the network) can
take the role of ``super-spreaders'' of the infection, depending on
the nature of the epidemic process and on the topological features of
the underlying network. In processes with no steady state, such as the
SIR model, the innermost core is the main trigger activating infection
and setting the value of the epidemic threshold. On the contrary, in
processes allowing an endemic steady-state, the actual activation
mechanism depends essentially on the degree of heterogeneity of the
network.  This simple picture, valid for uncorrelated networks, can be
however modified by the presence of strong degree correlations,
which can shift the weight towards one or the other mechanism.
These observations call for
further theoretical research on epidemic processes on strongly
correlated networks.  On the other hand, our results might find
practical applications in the implementation of optimized immunization
strategies, which can be designed to target the actual
``super-spreaders'' of a given contact network.  Moreover, our
  work could turn out to be relevant to other types of spreading
  processes, such as information, behavior or rumor spreading, and
  more in general, to dynamical processes whose behavior and possible
  transitions are ruled by the value of the largest eigenvalue of the
  adjacency matrix, such as, for example, synchronization phenomena. 
Also, for these latter dynamics, our results
call for additional investigations on the existence and interplay of
the relevant activation mechanisms.

\newpage 
\section*{Methods}

\subsection*{The SIS and SIR epidemic models}
In the SIS model, individuals can be in one of two states, either
susceptible or infected. Susceptibles become infected by contact with
infected individuals, with a rate equal to the number of infected
contacts times a given spreading rate $\lambda$. Infected individuals
on the other hand become healthy again with a rate $\mu$ that can be
taken arbitrarily equal to unity, thus setting the characteristic time
scale. This model allows thus individuals to contract the infection
time and again, making possible, in the infinite population limit, a
sustained infected steady state (endemic state). This occurs for
values of $\lambda$ larger than the epidemic threshold $\lambda_c$,
while for $\lambda<\lambda_c$ the epidemics lasts only for a finite
time and asymptotically all individuals are healthy.

\noindent In the SIR model, on the other hand, individuals can be in
one of three different states: susceptible, infected and recovered (or
removed). The dynamical rule for susceptible individuals is the same
as for SIS. With a rate $\mu$ (again set to unity) infected
individuals change their state and recover. Recovered individuals are
completely inert and cannot become infected again. With this dynamics
the system always reaches asymptotically an absorbing state with only
susceptible or removed individuals and no infected ones. A threshold
$\lambda_c$ separates a regime where outbreaks reach a finite fraction
of the individuals (i.e. the final density of removed individuals is
finite) from a regime $\lambda < \lambda_c$ where only an
infinitesimal fraction of individuals is hit.

\subsection*{The $k$-core decomposition.}
\begin{figure*}[h]
  \begin{center}
    \includegraphics[width=14cm,angle=0]{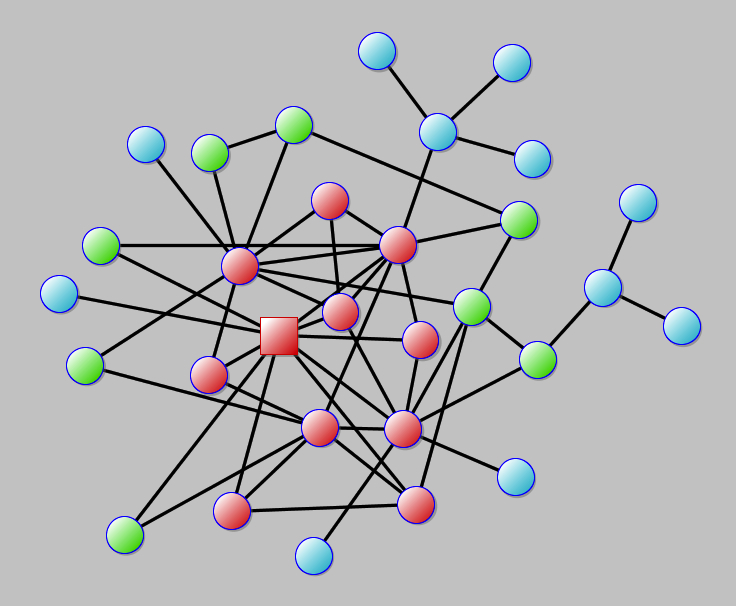}
  \end{center}
  \caption{\normalsize Visual representation of the $k$-core
      decomposition of a small network of size $N=30$ and maximum
      degree $q_{max}=10$.  Blue vertices belong to the $k=1$ shell
      and green vertices to the $k=2$ shell. The maximum $k$-core,
      with $k_S=3$, is composed by the red vertices. The hub (vertex
      with largest degree) is represented as a square.}
  \label{fig:examplenet}
\end{figure*}

\begin{figure*}[h]
  \begin{center}
    \includegraphics[width=16cm,angle=0]{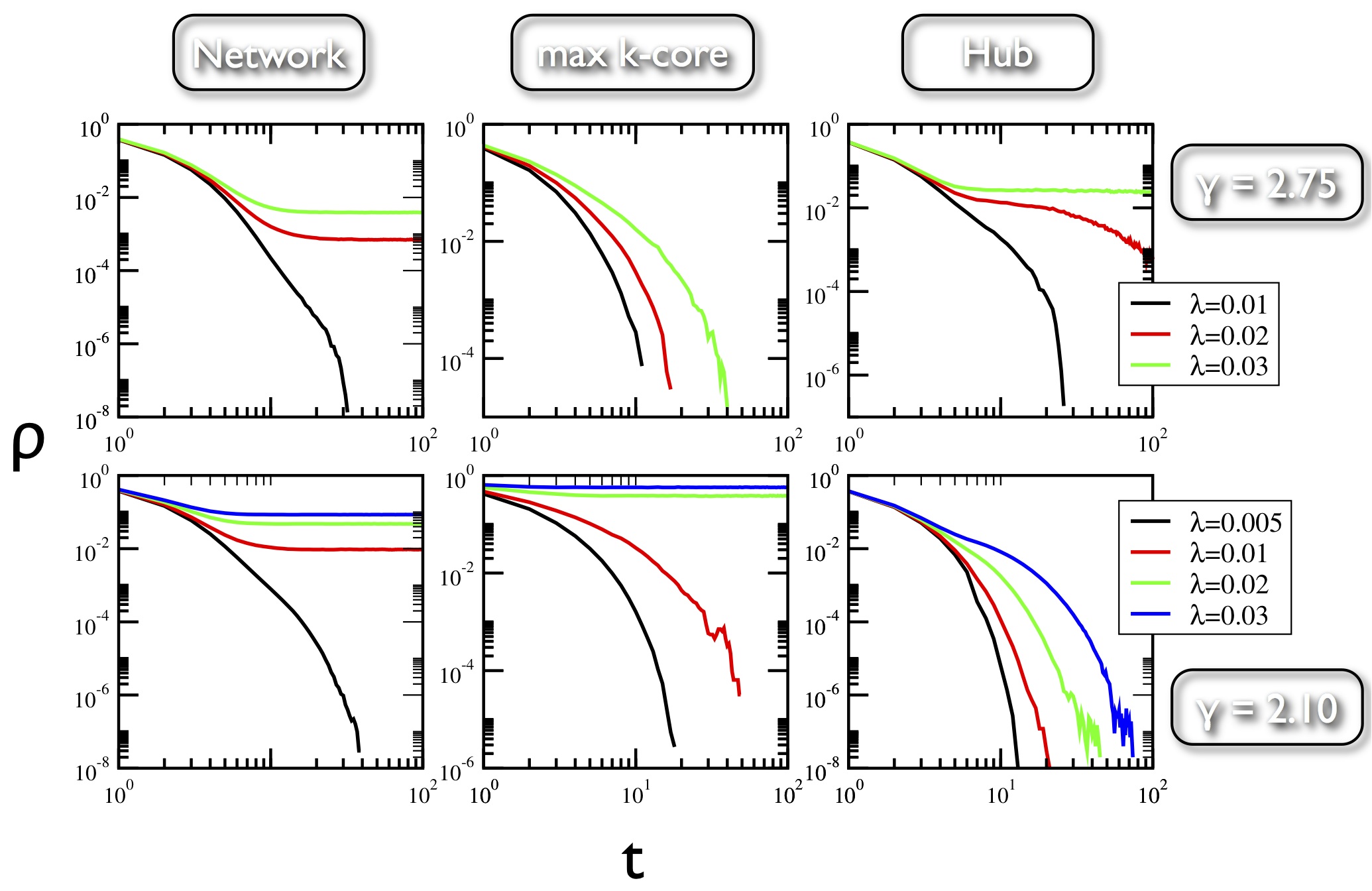}
  \end{center}
  \caption{\normalsize Average density of infected vertices as a
    function of time, $\rho(t)$, in the SIS model on uncorrelated
    scale-free networks generated by means of the UCM algorithm. We
    consider networks with widely separated degree exponent and
    different size, namely $\gamma = 2.75$, $N=3 \times 10^7$ and
    $\gamma=2.1$, $N=10^6$. The different columns correspond to the
    average density computed when the dynamics runs over the whole
    network (left), only over the maximum $k$-core of the network
    (center), and only over the largest hub (right), considered as an
    isolated star network. The different colors correspond to
    different values of the spreading rate $\lambda$. 
    For small $\gamma=2.1$ (bottom row), the onset of the global
    steady state is correlated with the active state of the epidemics
    on the maximum $k$-core, while it corresponds to a subcritical
    state for the hub. This observation indicates that in this case
    the maximum $k$-core is responsible for the overall activity in
    the network. For large $\gamma=2.75$, on the other hand, the
    global active state is linked to an active hub and a subcritical
    maximum $k$-core, signaling that it is the former mechanism the
    one keeping activity on a global scale.}
  \label{fig:UCMdata}
\end{figure*}

\begin{figure*}[h]
   \begin{center}
     \includegraphics[width=15cm,angle=0]{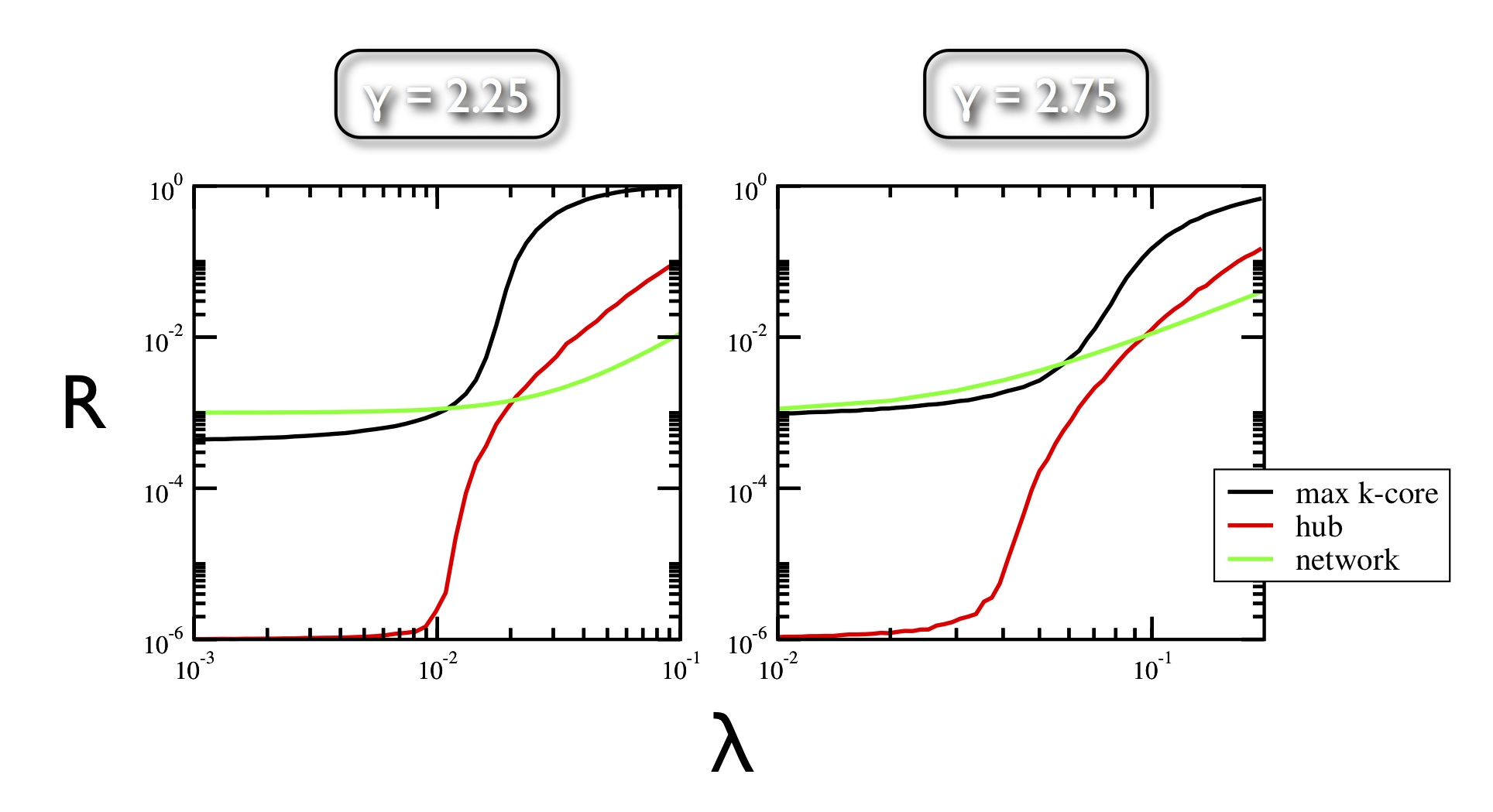}
   \end{center}
  \caption{\normalsize Total density $R$ of infected vertices, as a
    function of the spreading rate $\lambda$, computed after an
    epidemic outbreak in the SIR model on uncorrelated scale-free
    networks generated by means of the UCM algorithm. The degree
    exponents considered are $\gamma=2.25$ (left) and $\gamma=2.75$
    (right), with network sizes $N=10^6$. The line colors correspond
    to the SIR dynamics restricted to the maximum $k$-core (black), to
    the largest hub (green), and on the whole network (red). The value
    of $\lambda$ after which a macroscopic fraction of the network
    becomes infected is correlated in  the whole network and the
    maximum $k$-core, while the infection pattern on the hub conforms
    with the theoretical expression in equation~\eqref{eq:2}. These
    results indicate the crucial role of the maximum $k$-core in
    keeping the SIR activity on a large scale in networks,
    independently of the behavior of the hub.}
  \label{fig:UCMSIR}
\end{figure*}

\begin{figure*}[h]
   \begin{center}
     \includegraphics[width=16cm,angle=0]{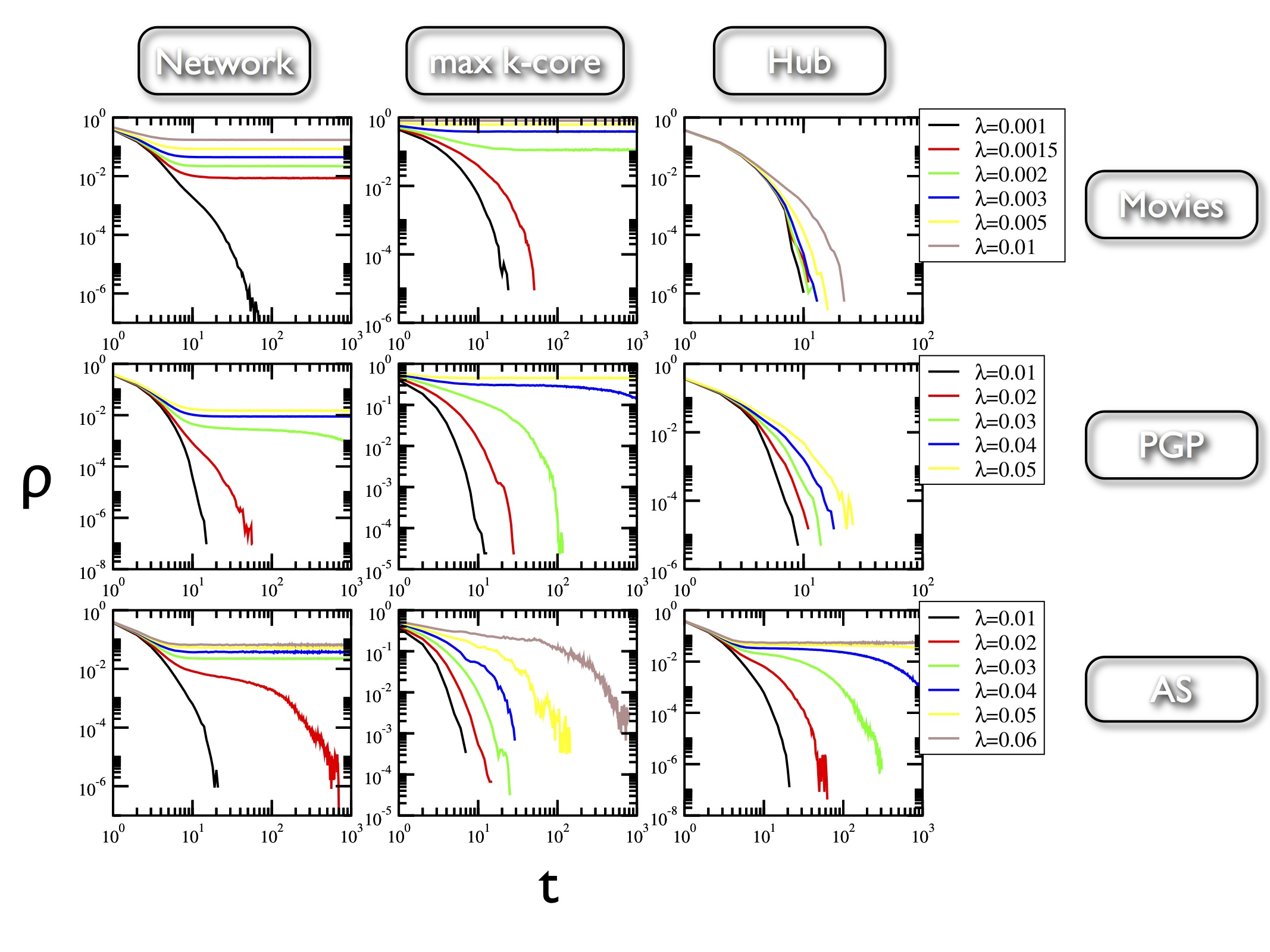}
   \end{center}
  \caption{\normalsize Average density of infected vertices as a
    function of time, $\rho(t)$, for the SIS model on three instances
    of real correlated networks: The network of actors co-starring in
    Hollywood movies (Movies), the social network of
    pretty-good-privacy (PGP), and an Internet map at the autonomous
    system level (AS); see Methods for further details of the
    networks. As in Fig.~\ref{fig:UCMdata}, columns refer to the
    activity on the whole network (left) and restricted to the maximum
    $k$-core (center) or the largest hub (right). Line colors indicate
    different values of the spreading rate considered in the networks.
    In the Movies and PGP networks, the maximum $k$-core dominates the
    transition, as expected due to the small degree exponent of all
    three networks ($\gamma \sim 2$). Surprisingly, in the AS network
    it is the activation of the hub the dominant mechanism setting in
    the steady state. This different behavior must be attributed to
    the very strong correlations present in the AS map (see main
    text).}
  \label{fig:Realdata}
\end{figure*}

The $k$-core decomposition is an iterative procedure to classify
vertices of a network in layers of increasing density of connections.
Starting with the full graph one removes the vertices with only one
connection (degree $q=1$). This procedure is then repeated until only
nodes with degree $q \ge 2$ are left. The removed nodes constitute the
$k=1$-shell and those remaining are the $k=2$-core. At the next step
all vertices with degree $q=2$ removed, thus leaving the $k=3$-core.
The procedure is repeated iteratively. The maximum $k$-core (of index
$k_S$) is the set of vertices such that one more iteration of the
procedure removes all of them. Notice that all vertices of the
$k$-core of index $k$ have degree larger than or equal to
$k$. Figure~\ref{fig:examplenet} shows an example of the
  $k$-core decomposition performed on a small network of size $N=30$
  and largest degree $q_{max}=10$.

\subsection*{SIR model on a star-graph}
Let us consider the SIR process on a star network of size $q_{max}
+1$. The process starts from a randomly infected vertex, and proceeds
till all infected vertices become eventually removed. We want to
compute the average final density of removed vertices at the end of an
outbreak, which is given by $R = r/ (q_{max}+1)$, where $r$ is the
total number of removed vertices. Let us define $r=1+r^*$, where $r^*$
is the number of removed vertices from secondary infections. The
outcome of the process will depend on whether the initial infected
site is the hub or a leaf (a vertex of degree $1$). If the infection
starts in the hub, which will happen with probability $p_h =
1/(q_{max}+1)$, the average number of secondary infected vertices will
be $\av{r^*}_h = \lambda q_{max}$. On the contrary, the infection can
start in a leaf with probability $p_l = q_{max}/(q_{max}+1)$. In this
case, the hub can become infected with probability $\lambda$, and from
there, spread the infection to the remaining $q_{max}-1$ susceptible
leaves. Therefore, in this case the average number of secondary
infections is $\av{r^*}_l = \lambda[1+\lambda (q_{max}-1)]$. The
average total number of removed sites at the end of the spreading
process will therefore be
\begin{eqnarray*}
	\av{r} &=& 1+ \av{r^*} = 1+ p_h \av{r^*}_h + p_l \av{r^*}_l 
	=
	1+\frac{\lambda q_{max}}{q_{max}+1} + \frac{q_{max}}{q_{max}+1}
	\lambda[1+\lambda (q_{max}-1)]\\
	&=& 1 + \frac{\lambda q_{max}}{q_{max}+1} [2+\lambda (q_{max}-1) ]
	\simeq 1 + 2 \lambda + \lambda^2 q_{max},
\end{eqnarray*}
where the last expression is valid in the limit of large $q_{max}$.
From here it follows the expression for the average final density of
infected vertices, equation~\eqref{eq:2}, valid also in the limit of
large $q_{max}$.

\subsection*{Quantitative features of the real networks considered}
We consider in our analysis the following three real networks
datasets:

\textbf{Internet map at the Autonomous System level (AS)}. Map of the
Internet collected at the Oregon route server. Vertices represent
autonomous systems (aggregations of Internet routers under the same
administrative policy), while edges represent the existence of border
gateway protocol (BGP) peer connections between the corresponding
autonomous systems \cite{romuvespibook}.

\textbf{Pretty-good-privacy network (PGP)}. Social network defined by
the users of the pretty-good-privacy (PGP) encryption algorithm for
secure information exchange. Vertices represent users of the PGP
algorithm. An edge between to vertices indicates that each user has
signed the encryption key of the other \cite{PhysRevE.70.056122}.

\textbf{Actor collaboration network (Movies)}. Network of movie actor
collaboration obtained from the Internet Movie Database (IMDB). Each
vertex represents a movie actor. Two actors are joined by an edge if
they have co-starred at least one movie \cite{Barabasi:1999}.

The relevant topological features of the different maps are summarized
in Table \ref{datasets}.

\section*{Acknowledgments}
RP-S acknowledges financial support from the Spanish MICINN, under
project No. FIS2010-21781-C02-01, and the Junta de Andaluc\'{\i}a,
under project No. P09-FQM4682, as well as additional support through
ICREA Academia, funded by the Generalitat de Catalunya.

\clearpage

\subsection*{Tables}

\begin{table*}[h]
  \begin{center}
    \begin{tabular}{||l|c|c|c|c||}
      \hline \hline
      & $N$ & $\av{q}$ & $q_{max}$ & $k_S$\\
      \hline      \hline 
      Movies & 81860 & 89.532 & 3789 & 359\\  \hline  
      PGP & 10680 & 4.554 & 205 & 31\\  \hline  
      AS & 11174 & 4.190 & 2389 & 17\\  \hline  
      \hline  
    \end{tabular}
  \end{center}
  \caption{\normalsize Topological features of the real network datasets
    considered. Network size $N$, average degree $\av{q}$, degree of
    the largest hub $q_{max}$, and index of the maximum $k$-core, $k_S$.}

  \label{datasets}
\end{table*}

\clearpage

\subsection*{Figure Captions}

\clearpage 

\section*{SUPPLEMENTARY MATERIAL}

\section{Details of SIS numerical simulations.}
The numerical simulations of SIS dynamics have been performed in the
standard rejection-free way~\cite{Spv01a}: At each time step one node
from the list of infected vertices is chosen at random and removed;
then, each of its neighbors is considered and, if healthy, infected
with probability $\lambda$.  Time is incremented by the inverse of the
total number of occupied nodes. Substrate networks are created using
the uncorrelated configuration model (UCM) \cite{Sucmmodel}.

Initially, all nodes are taken to be infected.
No variation is expected for any finite fraction of infected nodes
in the initial state. Spreading experiments, with a single infected
vertex in the initial state, are an interesting topic for future work.

Results are averaged over many realizations (at least 10)
of the dynamical process on a single
network. We have not averaged over different network samples, because
fluctuations in the various geometrical quantities (max k-core order,
max k-core size, hub connectivity) make averaged results less clear
(see Ref.~\cite{SPhysRevLett.105.218701}).

The system sizes considered for larger $\gamma$ are larger than for smaller
values of $\gamma$. This is motivated by the need to increase the
separation between $1/\sqrt{q_{max}}$ and $\av{q}/\av{q^2}$, in order
to allow for a clear distinction between the two mechanisms.

\section{Results for additional values of $\gamma$.}

\begin{figure}[p]
  \begin{center}
   \includegraphics[width=14cm,angle=0]{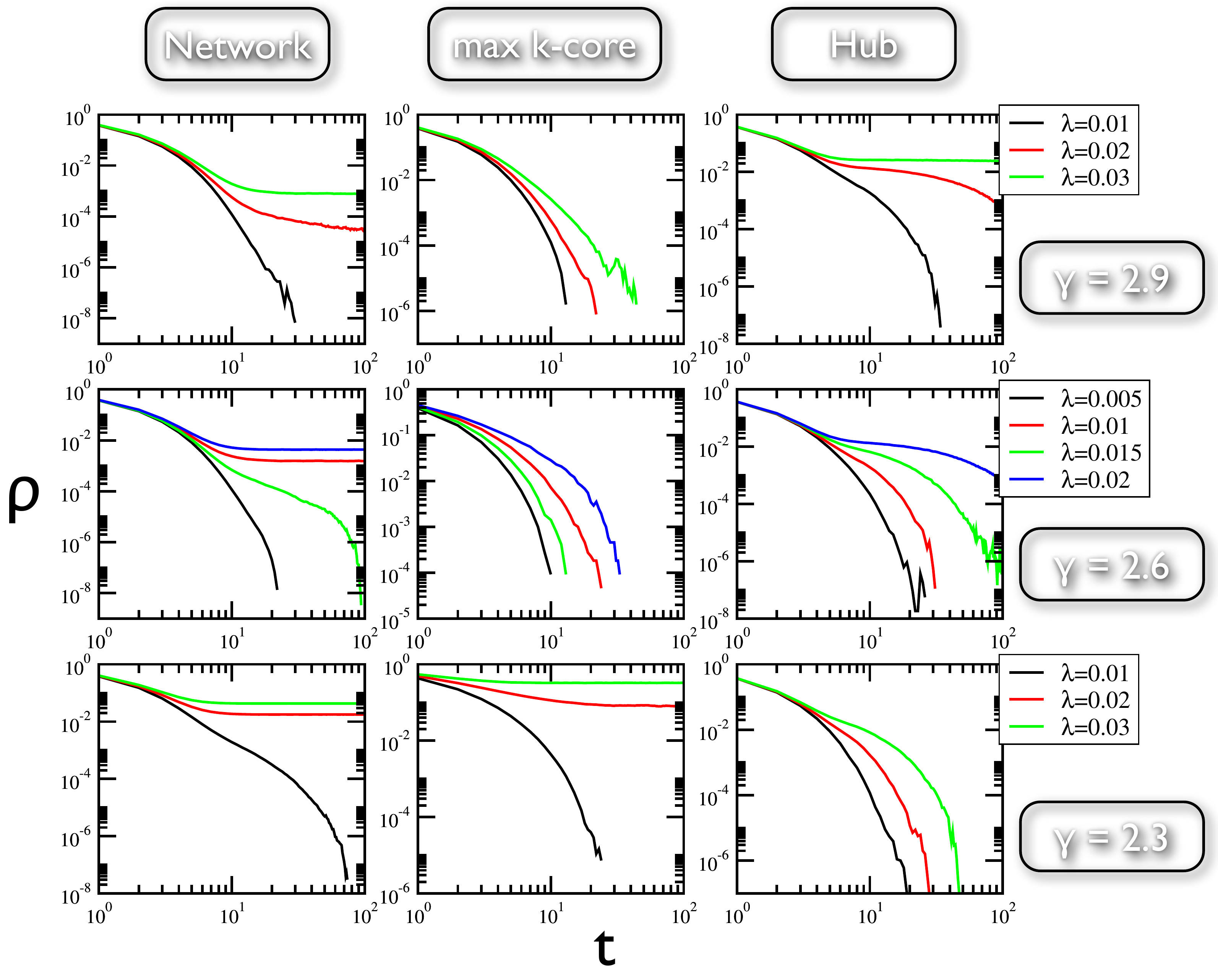}
 \end{center}
 \caption{\small Average density of infected vertices as a function of
   time, $\rho(t)$, in the SIS model on uncorrelated scale-free
   networks generated by means of the UCM algorithm.  We consider
   networks with $\gamma = 2.9$, $\gamma=2.6$ ($N=3 \times 10^7$) and
   $\gamma=2.3$ ($N=10^6$).  The different columns correspond to the
   average density computed when the dynamics runs over the whole
   network (left), only over the maximum $k$-core of the network
   (center), and only over the largest hub (right), considered as an
   isolated star network.  The different colors correspond to
   different values of the spreading rate $\lambda$. 
   For $\gamma=2.3$ (bottom row), the onset of the global steady state
   is correlated with the active state of the epidemics on the maximum
   $k$-core, while it corresponds to a subcritical state for the
   hub. This observation indicates that in this case the maximum
   $k$-core is responsible for the overall activity in the
   network. For large $\gamma>5/2$, on the other hand, the global
   active state is linked to an active hub and a subcritical maximum
   $k$-core, signaling that it is the former mechanism the one keeping
   activity on a global scale.  }
  \label{additionalSIS}
\end{figure}

In Fig.~\ref{additionalSIS} we report the results of additional
simulations of the SIS model on uncorrelated power-law distributed
networks with exponent $\gamma$, for values of $\gamma$ different from
those reported in the main text.  These results confirm that the
activation mechanism triggering the epidemic transition is the largest
hub or the innermost maximum $k$-core, depending on whether $\gamma$
is larger or smaller than $5/2$.

\section{Degree distribution of the maximum $k$-core.}

\begin{figure}[t]
  \begin{center}
    \includegraphics[width=8cm,angle=0]{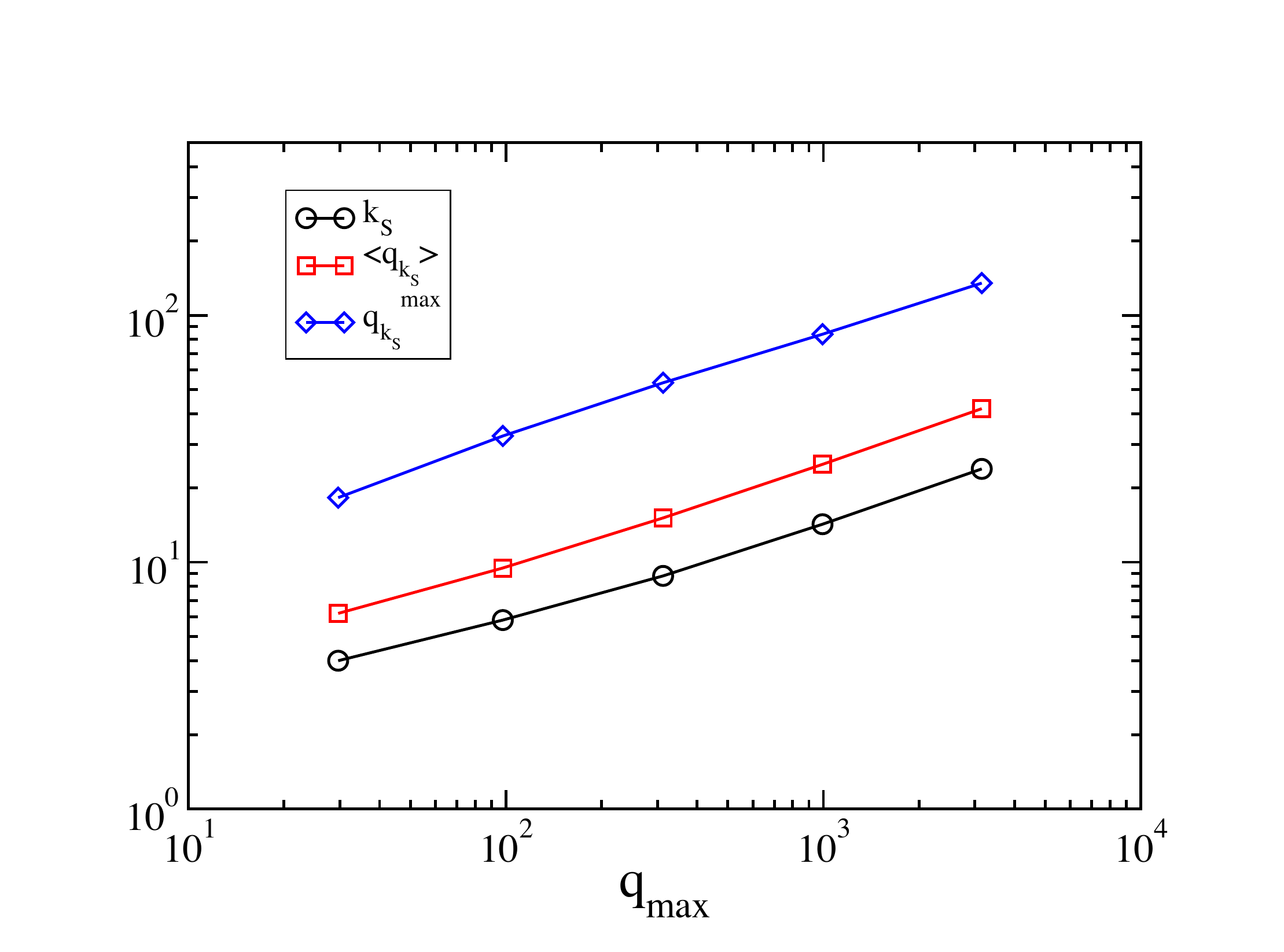}
 \end{center}
 \caption{\small Topological properties of the of the maximum $k$-core as a
   function of network size in uncorrelated scale-free networks with
   degree exponent $\gamma=2.5$.}
  \label{randomregulargraph}
\end{figure}

We study the topological properties of the maximum $k$-core in
uncorrelated scale-free networks generated using the UCM model. We
focus in particular on the average degree $\av{q_{k_S}}$ of the
maximum $k$-core, and on its maximum $q_{k_S}^{max}$ and minimum
$q_{k_S}^{min}=k_S$ degree. In Fig.~\ref{randomregulargraph} we plot
this quantities averaged over networks with different size $N$ as a
function of the network's maximum degree $q_{max} \sim N^{1/2}$. The
networks considered have a degree exponent $\gamma=2.5$. From this
figure, we conclude that all the computed properties of the maximum
$k$-core scale in the same way with the network size (i.e. are
proportional to each other), indicating that the degree distribution
remains rather narrow for any network size.  This observation
justifies the assumption that heterogeneous mean-field theory
determines correctly the scaling of the epidemic threshold on the
maximum $k$-core.

\section{Degree distributions and correlations in real networks.}

\begin{figure}[t]
  \begin{center}
    \includegraphics[width=8cm,angle=0]{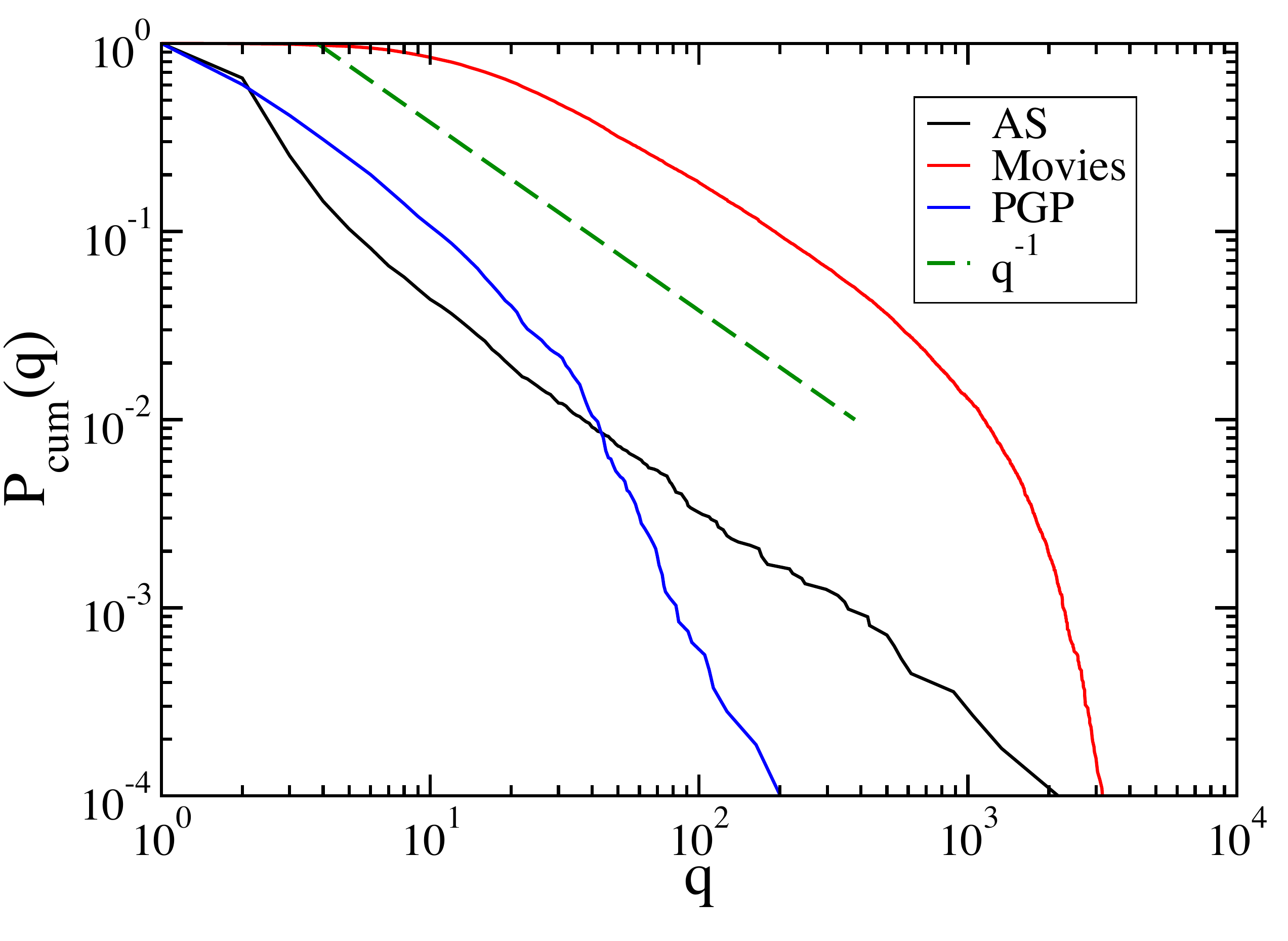}
 \end{center}
 \caption{\small Cumulated degree distribution of the real network
   datasets considered. All plots show a power-law regime compatible
   with the form $q^{-1}$, indicating a degree distribution with a
   degree exponent $\gamma \sim 2$.}
  \label{degreedistributions}
\end{figure}

In Fig.~\ref{degreedistributions} we report the degree distributions
of the three real networks considered. In all cases the distributions
decay slowly for large $q$ with an exponent close to $\gamma=2$.

\begin{figure}[t]
  \begin{center}
    \mbox{\includegraphics[width=7cm,angle=0]{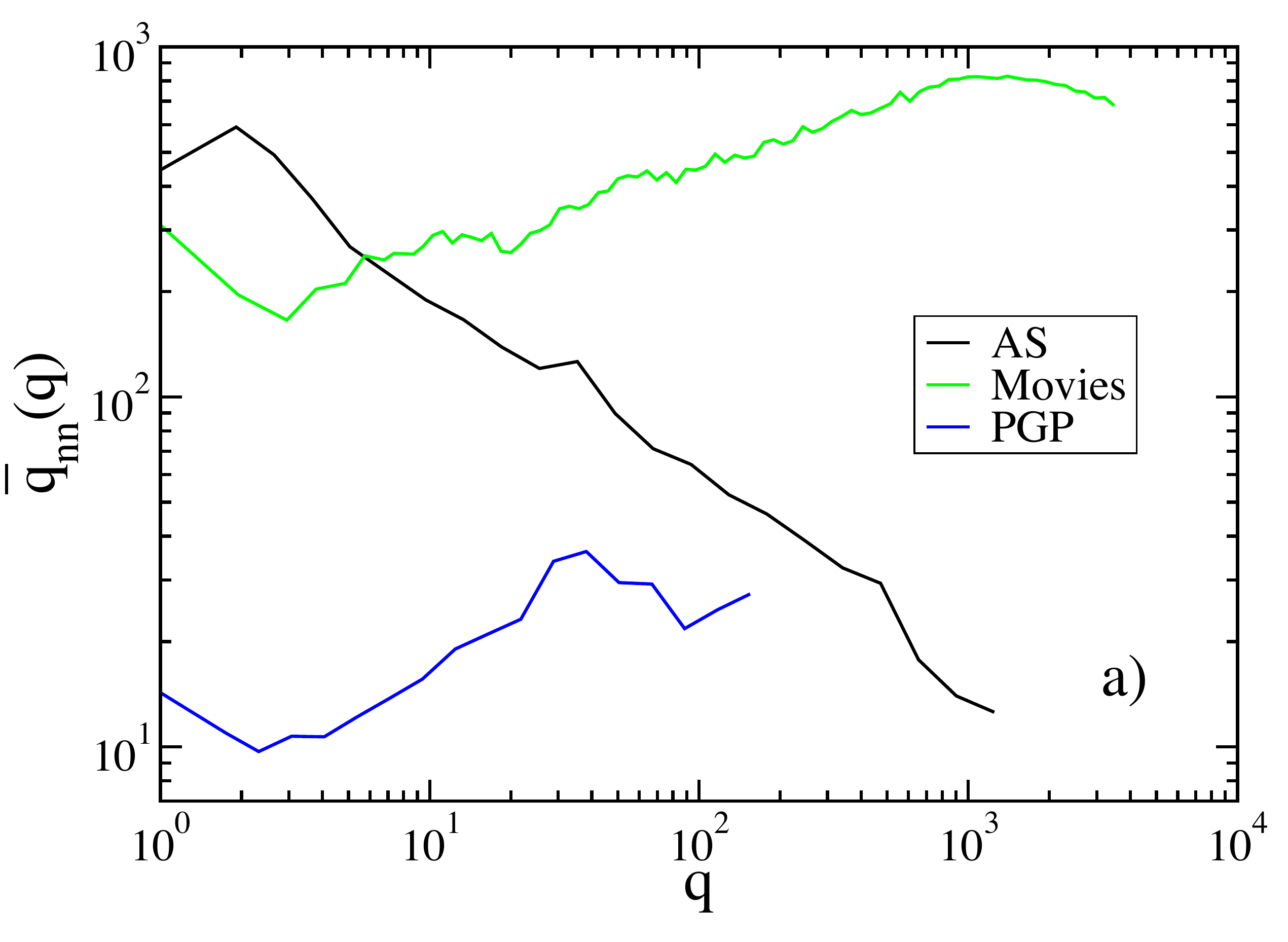}
    \includegraphics[width=7cm,angle=0]{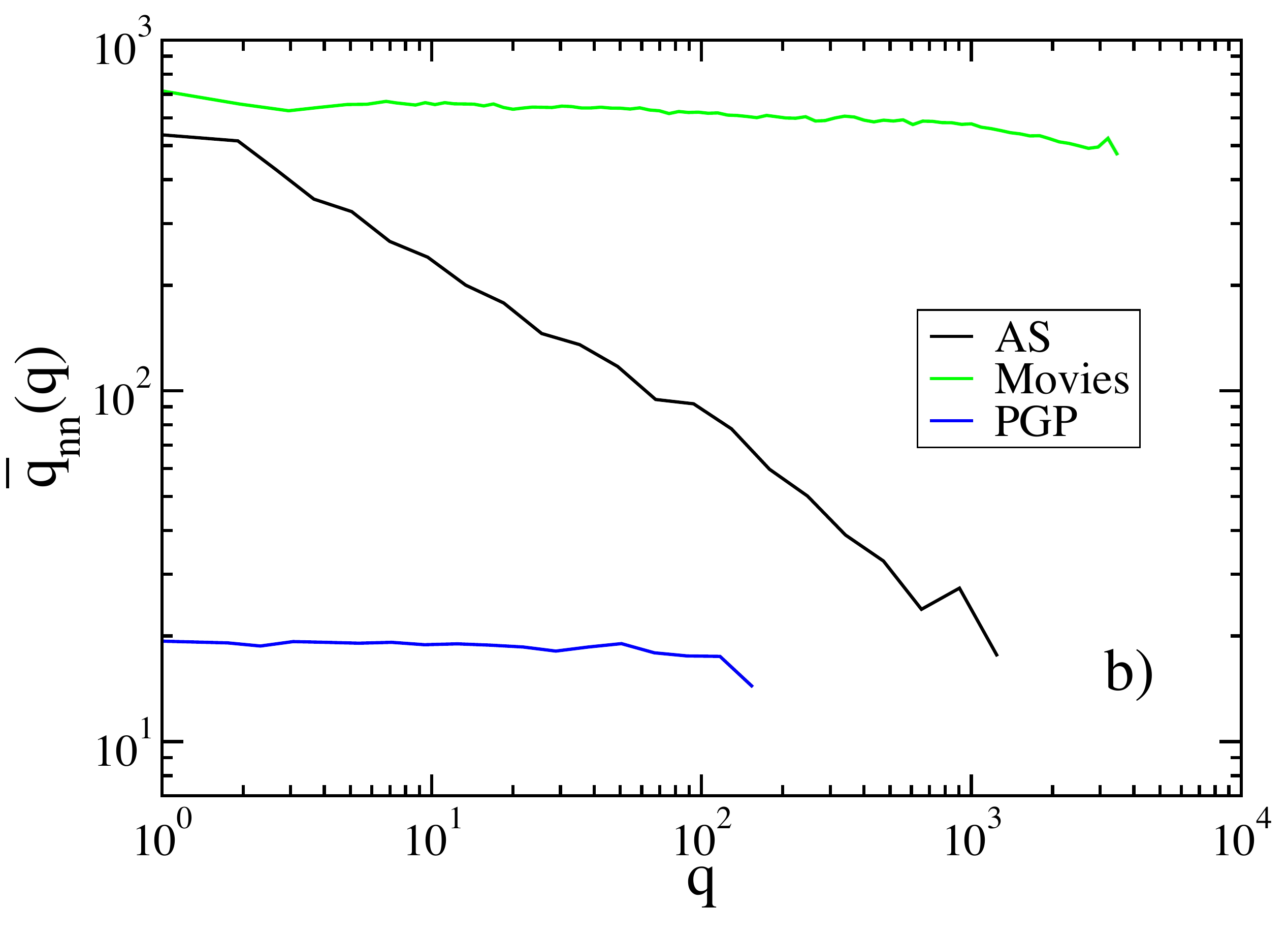}}
 \end{center}
  \caption{\small Average degree of the nearest neighbors (ANN) function,
    $\bar{q}_{nn}(q)$, for the three real networks considered. a)
    Original networks. b) Randomized networks, under application of
    the algorithm in Ref.~\cite{SMaslovSneppen}.}
  \label{knn}
\end{figure}

In Fig.~\ref{knn}(a) we plot on the other hand the average degree of the
nearest neighbors (ANN) of a node of degree $q$ for the three real
networks considered \cite{Salexei}.
The Movies and PGP networks exhibit a mild assortative structure
\cite{SPhysRevLett.89.208701}: vertices with high degree tend to be
connected in average with vertices of high degree, while low degree
vertices are connected to each other.  On the contrary, the AS network
is strongly disassortative, i.e.  vertices with large $q$ are
typically connected to nodes with small $q$, and viceversa.  The
different strength of degree correlations is highlighted by computing
the $\bar{q}_{nn}(q)$ function for the same networks after a
randomizing (degree-preserving) procedure is
applied~\cite{SMaslovSneppen}: two edges are randomly selected and two
vertices at their respective ends are swapped.

While the rewiring procedure completely destroys the weak assortative
correlations of the Movies and PGP networks, Fig.~\ref{knn}(b), no
effect can be spotted for the AS, indicating the presence of very
strong and robust correlations.

\section{Values of the thresholds} 

The numerical values of the analytical estimates of the different
thresholds discussed for the SIS in networks are reported in
Table~\ref{thresholdtable}. We focus in particular on the networks
considered in our manuscript, namely uncorrelated networks generated
with the UCM algorithm and three real network datasets.  The first
column shows the values of $\lambda_c^E=1/\Lambda_N$, the inverse of
the largest eigenvalue of the adjacency matrix.  In the second column,
we find the value of $1/\sqrt{q_{max}}$, which is the prediction for
uncorrelated networks with $\gamma>5/2$.  Finally, the third column
presents the value of $\langle q \rangle/\langle q^2 \rangle$, which
holds for $\gamma<5/2$ and coincides with the HMF prediction.  We
observe that for UCM networks the predictions of Eq.~(1) in the
manuscript are very well obeyed, indicating that the estimates of
Ref.~\cite{SChung03} are correct.  With regards to real networks, the
Movies network shows a threshold in good agreement with the prediction
of Eq.~(1) for $\gamma<5.2$, i.e. close to the HMF prediction,
indicating the relevant role of the maximum $k$-core.  For the PGP
network, the agreement with Eq.~(1) is not precise, yet the threshold
is closer to the HMF prediction and therefore the maximum $k$-core
drives the transition, as shown in Fig.~3 in the manuscript.  The
opposite is true for the AS network: the inverse of the largest
eigenvalue is close to $1/\sqrt{q_{max}}$ and is much larger than the
HMF prediction, despite a very slow decay of the degree
distribution. The conclusion is that in this case the estimate of
Ref.~\cite{SChung03} is largely incorrect, thus invalidating Eq.~(1).
This failure is originated by the disassortative correlations present
in AS networks, induced by the presence of a very large hub,
  which mixes the role of
  the maximum $k$-core and the hub in the activation of the epidemics.

\begin{table}[h]
  \begin{center}
    \begin{tabular}{||l|c|c|c||}
      \hline \hline
      & $1/\Lambda_N$ & $1/\sqrt{q_{max}}$ & $\av{q}/\av{q^2}$ \\
      \hline
      & \multicolumn{3}{c||}{UCM model} \\ \hline
      $\gamma=2.9$, $N=3 \cdot 10^7$   & 0.01352 & 0.01356 & 0.03896 \\  \hline
      $\gamma=2.75$, $N=3 \cdot 10^7$  & 0.01308 & 0.01353 & 0.02312 \\  \hline
      $\gamma=2.6$, $N=3 \cdot 10^7$   & 0.01031 & 0.01351 & 0.01294 \\  \hline
      $\gamma=2.3$,  $N=10^6$          & 0.01160 & 0.03168 & 0.01209 \\  \hline
      $\gamma=2.1$,  $N=10^6$          & 0.00735 & 0.03163 & 0.00745 \\ 
      \hline \hline
      & \multicolumn{3}{c||}{Real networks} \\ \hline
      Movies & 0.001223& 0.01624 & 0.00168  \\ \hline
      PGP & 0.02356 & 0.06984 & 0.05296  \\ \hline
      AS & 0.016576 & 0.02045	& 0.003765  \\ \hline
      \hline  
    \end{tabular}
  \end{center}
  \caption{\small Table of SIS thresholds for synthetic and real networks.}
  \label{thresholdtable}
\end{table}

\begin{figure}[p]
  \begin{center}
   \includegraphics[width=17cm,angle=0]{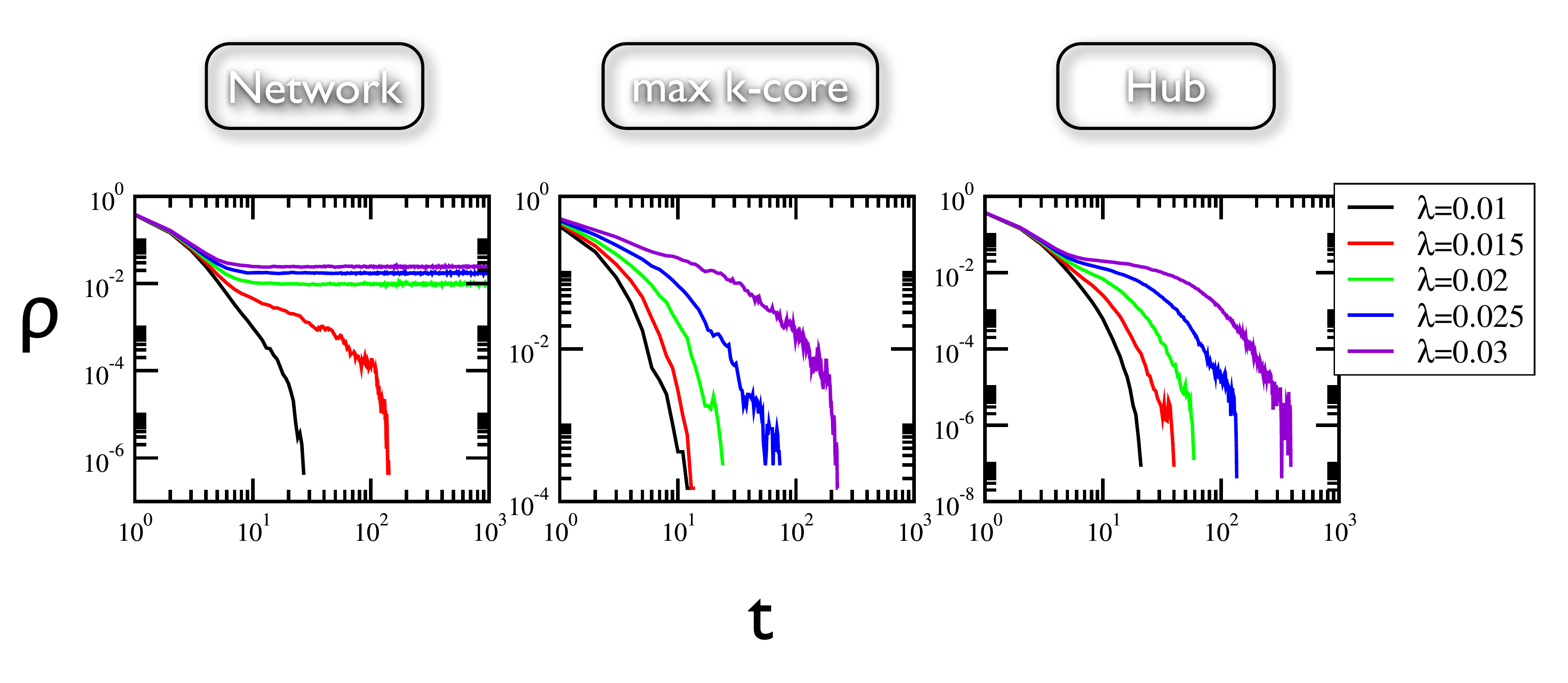}
 \end{center}
 \caption{\small Average density of infected vertices as a function of
   time, $\rho(t)$, in the SIS model on the  AS2 Internet map.  The
   different columns correspond to the average density computed when
   the dynamics runs over the whole network (left), only over the
   maximum $k$-core of the network (center), and only over the largest
   hub (right), considered as an isolated star network.  The different
   colors correspond to different values of the spreading rate
   $\lambda$. }
  \label{additionalAS}
\end{figure}

To further check the role of correlations and a large hub, we have
considered an additional AS map, corresponding to later snapshot of
the Internet, and thus characterized by a larger size. The
characteristics of this second Internet map (AS2) are the following:
$N=24463$, $\av{q} = 4.2799$, $k_S = 24$, $1/\Lambda_N = 0.01412$
$1/\sqrt{q_{max}} = 0.02012$ $\av{q}/\av{q^2} = 0.003796$.  In
Figure~\ref{additionalAS} we present the results of numerical
simulations of the SIS model performed on this AS2 map. We observe
that, in this case, for values of $\lambda$ corresponding to an active
global network, neither the hub nor the maximum $k$-core are
apparently in the active phase. This result, which is also found in
other AS maps we have considered, confirms that the presence of very
strong correlations alters the clear picture observed in uncorrelated
networks, leading to a mixing of the activation mechanisms.

\section{SIS dynamics on Weber-Porto networks.}

\begin{figure}
  \begin{center}
    \includegraphics[width=18cm,angle=0]{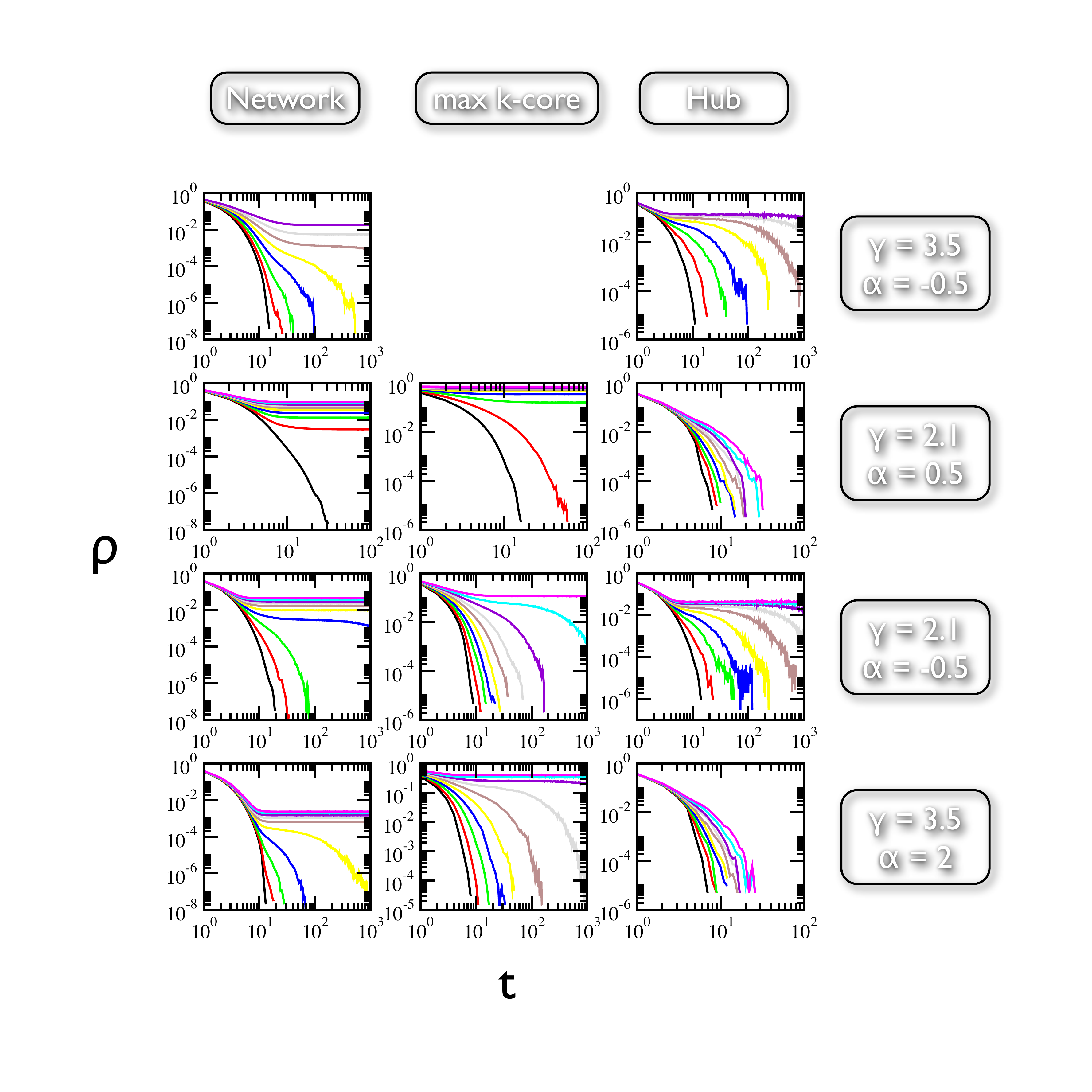}
  \end{center}
\vspace*{-2cm}
\caption{\small Density of infected vertices as a
  function of time in the SIS model on Weber-Porto networks with
  different values of $\gamma$ and $\alpha$. Network sizes $N=10^4$.
  We plot the average density over the whole network (left column),
  over the maximum $k$-core (center column), and over the largest hub
  (right columns). The different colors correspond, from bottom to top
  of each plot, to increasing values of the spreading rate $\lambda$.}
  \label{WeberPortodata}
\end{figure}

Recently, Weber and Porto~\cite{SPhysRevE.76.046111} have introduced a
modified configuration model \cite{Smolloy95} which allows to generate
graphs with a given degree distribution and, simultaneously, a
predetermined ANN function $\bar{q}_{nn}(q)$.  We have performed
simulations of the SIS dynamics on networks generated by the
Weber-Porto method, and compared the density of active nodes as a
function of $\lambda$ when the dynamics takes place on the whole
network or only on the maximum $k$-core or on a star-graph with
$q_{max}+1$ nodes, where $q_{max}$ is the degree of the largest hub.
Numerical values of the different theoretical thresholds, for
different values of the degree exponent $\gamma$ and of the exponent
$\alpha$ of the ANN function, $\bar{q}_{nn}(q) \sim q^{\alpha}$, are
reported in Table~\ref{thresholdtable2}.  Positive values of $\alpha$
imply assortative correlations, while $\alpha<0$ is the signature of
disassortative correlations.
Results of numerical simulations are presented in Fig.~\ref{WeberPortodata}.

\begin{table}[h]
  \begin{center}
    \begin{tabular}{||l|c|c|c||}
      \hline \hline
      & $1/\Lambda_N$ & $1/\sqrt{q_{max}}$ &
      $\av{q}/\av{q^2}$ \\
      \hline
      \hline 
      & \multicolumn{3}{c||}{Weber-Porto network model} \\ \hline
      $\gamma=2.1$, $\alpha=0.5$   & 0.008548 & 0.05634 & 0.01798  \\  \hline
      $\gamma=3.5$, $\alpha=-0.5$  & 0.064435 & 0.06482 & 0.14492  \\  \hline
      $\gamma=2.1$, $\alpha=-0.5$  & 0.016137 & 0.01754 & 0.00353  \\  \hline
      $\gamma=3.5$, $\alpha=2$     & 0.021856 & 0.07001 & 0.14675  \\  \hline
      \hline  
    \end{tabular}
    \caption{\small Table of SIS thresholds for Weber-Porto networks.}
    \label{thresholdtable2}
  \end{center}
\end{table}

In the first two cases presented in Table~\ref{thresholdtable2},
correlations are expected to strengthen the mechanism already driving
the transition for uncorrelated networks: indeed for $\gamma>5/2$
disassortative correlations make the network more ``star-like'', while
for $\gamma<5/2$ assortative correlations are expected to make the max
$k$-core play an even more important role.  This is confirmed by the
data in Table~\ref{thresholdtable2}:
\begin{itemize}
\item For $\gamma=2.1$ and $\alpha=0.5$, the threshold $1/\Lambda_N$
is smaller than the HMF prediction (which is in its turn smaller than
  $1/\sqrt{q_{max}}$).
\item For $\gamma=3.5$ and $\alpha=-0.5$ the threshold $1/\Lambda_N$
is in very good agreement with $1/\sqrt{q_{max}}$.
\end{itemize}
Eq.~(1) in the manuscript turns out to be essentially correct.
Numerical results in Supplementary Figure~\ref{WeberPortodata}
confirm these findings. Notice that for $\gamma=3.5$ and $\alpha=-0.5$
there is no $k$-core structure in the network.
Numerical results (Fig.~\ref{WeberPortodata}) confirm these findings. 

In the other two cases, correlations act against the mechanism
at work on uncorrelated networks.
Are they able to completely perturb the picture and change the
activation mechanism of the epidemics?
\begin{itemize}
\item
For $\gamma=2.1$ and disassortative networks ($\alpha=-0.5$), the threshold $1/\Lambda_N$
is larger than the HMF prediction and it coincides with $1/\sqrt{q_{max}}$.
This indicates that the largest hub drives the global transition,
at odds with the prediction of Eq.~(1).
Here correlations change completely the behavior, leading to a violation
of the estimate of $\Lambda_N$ on Ref.~\cite{SChung03}. The direct numerical simulations, presented in
Figure~\ref{WeberPortodata}, confirm these findings.
\item
For $\gamma=3.5$ and assortative correlations ($\alpha=2$), the threshold
$1/\Lambda_N$ is much smaller than $1/\sqrt{q_{max}}$
(which is in its turn smaller than the HMF prediction).
This shows again that the result of Ref.~\cite{SChung03} does not hold.
The direct numerical simulations,
Figure~\ref{WeberPortodata}, clearly shows that it is the
maximum $k$-core which triggers the transition in this case.
\end{itemize}


\begin{thebibliography}{10}

\bibitem{PhysRevLett.91.247901}
Cohen, R., Havlin, S.  \& ben Avraham, D.
\newblock Efficient immunization strategies for computer networks and
  populations.
\newblock {\em Phys. Rev. Lett.}{ \bf 91}, 247901 (2003).

\bibitem{Leskovec:2007:DVM:1232722.1232727}
Leskovec, J., Adamic, L.~A.  \& Huberman, B.~A.
\newblock The dynamics of viral marketing.
\newblock {\em ACM Trans. Web}{ \bf 1} May  (2007).

\bibitem{Daley64}
Daley, D.~J. \& Kendall, D.~G.
\newblock Epidemics and rumours.
\newblock {\em Nature}{ \bf 204}, 1118 (1964).

\bibitem{barabasi02}
Albert, R. \& Barab\'asi, A.-L.
\newblock Statistical mechanics of complex networks.
\newblock {\em Rev. Mod. Phys.}{ \bf 74}, 47--97 (2002).

\bibitem{Dorogovtsev:2002}
Dorogovtsev, S.~N. \& Mendes, J. F.~F.
\newblock Evolution of networks.
\newblock {\em Advances in Physics}{ \bf 51}, 1079--1187 (2002).

\bibitem{newman2003saf}
Newman, M.
\newblock {The Structure and Function of Complex Networks}.
\newblock {\em SIAM Review}{ \bf 45}, 167 (2003).

\bibitem{keeling05:_networ}
Keeling, M.~J. \& Eames, K. T.~D.
\newblock Networks and epidemic models.
\newblock {\em J. R. Soc. Interface}{ \bf 2}, 295--307 (2005).

\bibitem{anderson92}
Anderson, R.~M. \& May, R.~M., {\em Infectious diseases in humans}, (Oxford
  University Press, Oxford, 1992).

\bibitem{barratbook}
Barrat, A., Barth\'{e}lemy, M.  \& Vespignani, A., {\em Dynamical Processes on
  Complex Networks}, (Cambridge University Press, Cambridge, 2008).

\bibitem{serrano07:_correl}
Serrano, M.~A., Bogu{\~n}{\'a}, M., Pastor-Satorras, R.  \& Vespignani, A.
\newblock Correlations in complex networks.
\newblock In {\em Large scale structure and dynamics of complex networks: From
  information technology to finance and natural sciences, }Caldarelli, G. \&
  Vespignani, A., editors,  35--66 (World Scientific, Singapore, 2007).

\bibitem{marianproc}
Bogu{\~n}{\'a}, M., Pastor-Satorras, R.  \& Vespignani, A.
\newblock Epidemic spreading in complex networks with degree correlations.
\newblock In {\em Statistical Mechanics of Complex Networks, }Pastor-Satorras,
  R., Rub{\'\i}, J.~M.  \& D{\'\i}az-Guilera, A., editors, volume 625 of {\em
  Lecture Notes in Physics}. Springer Verlag, Berlin (2003).

\bibitem{Lloyd18052001}
Lloyd, A.~L. \& May, R.~M.
\newblock How viruses spread among computers and people.
\newblock {\em Science}{ \bf 292}, 1316--1317 (2001).

\bibitem{PhysRevE.66.016128}
Newman, M. E.~J.
\newblock Spread of epidemic disease on networks.
\newblock {\em Phys. Rev. E}{ \bf 66}, 016128 (2002).

\bibitem{pv01a}
Pastor-Satorras, R. \& Vespignani, A.
\newblock Epidemic spreading in scale-free networks.
\newblock {\em Phys. Rev. Lett.}{ \bf 86}, 3200--3203 (2001).

\bibitem{PhysRevLett.90.028701}
Bogu\~n\'a, M., Pastor-Satorras, R.  \& Vespignani, A.
\newblock Absence of epidemic threshold in scale-free networks with degree
  correlations.
\newblock {\em Phys. Rev. Lett.}{ \bf 90}, 028701 (2003).

\bibitem{Wang03}
Wang, Y., Chakrabarti, D., Wang, C.  \& Faloutsos, C.
\newblock Epidemic spreading in real networks: An eigenvalue viewpoint.
\newblock In {\em 22nd International Symposium on Reliable Distributed Systems
  (SRDS'03)},  25--34. IEEE Computer Society, Los Alamitos, CA, USA (2003).

\bibitem{0295-5075-89-3-38009}
G{\'o}mez, S., Arenas, A., Borge-Holthoefer, J., Meloni, S.  \& Moreno, Y.
\newblock Discrete-time markov chain approach to contact-based disease
  spreading in complex networks.
\newblock {\em Europhys. Lett.}{ \bf 89}(3), 38009 (2010).

\bibitem{Mieghem2011}
{Van Mieghem}, P.
\newblock The {N}-intertwined {SIS} epidemic network model.
\newblock {\em Computing}{ \bf 93}, 147---169 (2011).

\bibitem{Chung03}
Chung, F., Lu, L.  \& Vu, V.
\newblock Spectra of random graphs with given expected degrees.
\newblock {\em Proc. Natl. Acad. Sci. USA}{ \bf 100}, 6313--6318 (2003).

\bibitem{PhysRevLett.105.218701}
Castellano, C. \& Pastor-Satorras, R.
\newblock Thresholds for epidemic spreading in networks.
\newblock {\em Phys. Rev. Lett.}{ \bf 105}, 218701 (2010).

\bibitem{Ganesh05}
Ganesh, A., Massouli\'{e}, L.  \& Towsley, D.
\newblock The effect of network topology on the spread of epidemics.
\newblock In {\em IEEE INFOCOM},  1455--1466,  (2005).

\bibitem{Chatterjee09}
Chatterjee, S. \& Durrett, R.
\newblock Contact processes on random graphs with power law degree distributions have critical value 0.
\newblock {\em Annals of Probability}{ \bf 37}, 2332--2356 (2009).

\bibitem{Durrett10}
Durrett, R.
\newblock Some features of the spread of epidemics and information on a r andom
  graph.
\newblock {\em Proc. Natl. Acad. Sci. USA}{ \bf 107}, 4491--4498 (2010).

\bibitem{kitsak2010}
Kitsak, M., Gallos, L.~K., Havlin, S., Liljeros, F., Muchnik, L., Stanley,
  H.~E.  \& Makse, H.~A.
\newblock Identification of influential spreaders in complex networks.
\newblock {\em Nat. Phys.}{ \bf 6}, 888--893 (2010).

\bibitem{Seidman1983269}
Seidman, S.~B.
\newblock Network structure and minimum degree.
\newblock {\em Social Networks}{ \bf 5}, 269 --287 (1983).

\bibitem{PhysRevE.65.036104}
Pastor-Satorras, R. \& Vespignani, A.
\newblock Immunization of complex networks.
\newblock {\em Phys. Rev. E}{ \bf 65}(3), 036104 Feb  (2002).

\bibitem{Centola2010Spreading}
Centola, D.
\newblock The spread of behavior in an online social network experiment.
\newblock {\em Science}{ \bf 329}(5996), 1194--1197 09  (2010).

\bibitem{1367-2630-13-12-123005}
L{\"u}, L., Chen, D.-B.  \& Zhou, T.
\newblock The small world yields the most effective information spreading.
\newblock {\em New Journal of Physics}{ \bf 13}(12), 123005 (2011).

\bibitem{Arenas200893}
Arenas, A., D{\'\i}az-Guilera, A., Kurths, J., Moreno, Y.  \& Zhou, C.
\newblock Synchronization in complex networks.
\newblock {\em Physics Reports}{ \bf 469}(3), 93 -- 153 (2008).

\bibitem{ucmmodel}
Catanzaro, M., {Bogu\~{n}\'{a}}, M.  \& Pastor-Satorras, R.
\newblock Generation of uncorrelated random scale-free networks.
\newblock {\em Phys. Rev. E}{ \bf 71}, 027103 (2005).

\bibitem{PhysRevLett.96.040601}
Dorogovtsev, S.~N., Goltsev, A.~V.  \& Mendes, J. F.~F.
\newblock $k$-core organization of complex networks.
\newblock {\em Phys. Rev. Lett.}{ \bf 96}, 040601 (2006).

\bibitem{Cohen00}
Cohen, R., Erez, K., ben Avraham, D.  \& Havlin, S.
\newblock Resilience of the internet to random breakdowns.
\newblock {\em Phys. Rev. Lett.}{ \bf 85}(21), 4626--4628 Nov  (2000).

\bibitem{newman02}
Newman, M. E.~J.
\newblock The spread of epidemic disease on networks.
\newblock {\em Phys. Rev. E}{ \bf 66}, 016128 (2002).

\bibitem{PhysRevLett.89.208701}
Newman, M. E.~J.
\newblock Assortative mixing in networks.
\newblock {\em Phys. Rev. Lett.}{ \bf 89}, 208701 (2002).

\bibitem{alexei}
Pastor-Satorras, R., V{\'a}zquez, A.  \& Vespignani, A.
\newblock Dynamical and correlation properties of the {I}nternet.
\newblock {\em Phys. Rev. Lett.}{ \bf 87}, 258701 (2001).

\bibitem{PhysRevE.76.046111}
Weber, S. \& Porto, M.
\newblock Generation of arbitrarily two-point-correlated random networks.
\newblock {\em Phys. Rev. E}{ \bf 76}, 046111 (2007).

\bibitem{romuvespibook}
Pastor-Satorras, R. \& Vespignani, A., {\em Evolution and structure of the
  Internet: A statistical physics approach}, (Cambridge University Press,
  Cambridge, 2004).

\bibitem{PhysRevE.70.056122}
Bogu\~n\'a, M., Pastor-Satorras, R., D\'\i{}az-Guilera, A.  \& Arenas, A.
\newblock Models of social networks based on social distance attachment.
\newblock {\em Phys. Rev. E}{ \bf 70}, 056122 (2004).

\bibitem{Barabasi:1999}
Barab{\'a}si, A.-L. \& Albert, R.
\newblock Emergence of scaling in random networks.
\newblock {\em Science}{ \bf 286}, 509--512 (1999).

\end{thebibliography}

\clearpage

\begin{thebibliography}{1}

\bibitem{Spv01a}
Pastor-Satorras, R. \& Vespignani, A.
\newblock Epidemic spreading in scale-free networks.
\newblock {\em Phys. Rev. Lett.}{ \bf 86}, 3200--3203 (2001).

\bibitem{Sucmmodel}
Catanzaro, M., {Bogu\~{n}\'{a}}, M.  \& Pastor-Satorras, R.
\newblock Generation of uncorrelated random scale-free networks.
\newblock {\em Phys. Rev. E}{ \bf 71}, 027103 (2005).

\bibitem{SPhysRevLett.105.218701}
Castellano, C. \& Pastor-Satorras, R.
\newblock Thresholds for epidemic spreading in networks.
\newblock {\em Phys. Rev. Lett.}{ \bf 105}, 218701 (2010).

\bibitem{Salexei}
Pastor-Satorras, R., V{\'a}zquez, A.  \& Vespignani, A.
\newblock Dynamical and correlation properties of the {I}nternet.
\newblock {\em Phys. Rev. Lett.}{ \bf 87}, 258701 (2001).

\bibitem{SPhysRevLett.89.208701}
Newman, M. E.~J.
\newblock Assortative mixing in networks.
\newblock {\em Phys. Rev. Lett.}{ \bf 89}, 208701 (2002).

\bibitem{SMaslovSneppen}
Maslov, S. \& Sneppen, K.
\newblock Specificity and stability in topology of protein networks.
\newblock {\em Science}{ \bf 296}, 910--913 (2002).

\bibitem{SChung03}
Chung, F., Lu, L.  \& Vu, V.
\newblock Spectra of random graphs with given expected degrees.
\newblock {\em Proc. Natl. Acad. Sci. USA}{ \bf 100}, 6313--6318 (2003).

\bibitem{SPhysRevE.76.046111}
Weber, S. \& Porto, M.
\newblock Generation of arbitrarily two-point-correlated random networks.
\newblock {\em Phys. Rev. E}{ \bf 76}, 046111 (2007).

\bibitem{Smolloy95}
Molloy, M. \& Reed, B.
\newblock A critical point for random graphs with a given degree sequence.
\newblock {\em Random Struct. Algorithms}{ \bf 6}, 161 (1995).

\end{thebibliography}
\end{document}